\documentclass{article}
\usepackage{multicol}
\usepackage{txfonts}
\usepackage{float}
\usepackage[dvipdfm]{graphicx}

\begin{document}

\title{\bf Spectroscopy and Multi-color Photometry of \\U Scorpii at the Earliest Stage of 2010 Outburst}

\author{
        Kazuyoshi IMAMURA and Kenji TANABE\\ \\
        \small \it Graduate School of Informatics, Okayama University of Science,\\ 
        \small \it 1-1 Ridai-cho, Kita-ku, Okayama, 700-0005, JAPAN
        }
        
\date{\small } 

\maketitle
\thispagestyle{empty}

\begin{abstract} 

\small \bf We have performed our spectroscopic and multi-color photometric observations of 
the recurrent nova U Scorpii at the earliest stage of the 2010 outburst. 
0.37 days after the discovery of the outburst, we can see broad and prominent emission lines of Balmer series, 
He I, N II, N III, O I and Mg II on the spectra. 
The FWHM of H$\alpha$ line yields an expansion velocity of approximately 6200 km/s. 
This line also accompanies a blue shifted absorption line (so called P Cygni profile). 
1.37 days after the discovery, H$\alpha$ line shows a nearly flat-topped profile in contrast to the previous day. 
From our multi-color photometry, we can see its rapid decline (one magnitude per day) of the brightness in each color band. 

\end{abstract}

\begin{flushleft} 
\end{flushleft}

\vspace*{0.2cm}
\begin{multicols}{2}


\section*{\normalsize 1. Introduction}
\hspace{0.5cm} U Scorpii (we denote U Sco, hereafter) is a recurrent nova whose outburst occurs every 8-12 years. Previous outbursts were recorded in 1863, 1906, 1917, 1936, 1945, 1969, 1979, 1987 and 1999 (Schaefer, 2010 $^1$). The first detection was by N. R. Pogson in 1863 $^2$. They were all characterized by a very fast evolution. U Sco is also known as a very fast nova defined by 
$t_2 < 10$ d, where $t_2$ is an elapsed time when the object faded by 2 magnitude from the maximum (Payne-Gaposchkin, 1957 $^3$).

  The tenth outburst of the recurrent nova U Sco was detected by S. Dvorak and B. G. Harris independently on January 28.4743UT and 28.4385UT 2010, respectively. Immediately after receiving these announcements (Maehara, 2010 $^4$ and Schaefer et al., 2010 $^5$), we started our spectroscopic and multi-color photometric observations at the OUS (Okayama University of Science) observatory simultaneously (on Jan. 28.84UT and Jan. 29.84UT).
  
  The purpose of this paper is to report the details of the earliest observations of U Sco in its 2010 outburst. This report will serve the comprehensive research and future understanding of this object.

\section*{\normalsize 2. Observation}

\hspace{0.5cm}  The first night observation is at 0.37 days after discovery. This is probably the earliest observation because of geographical reason. Our spectroscopic observational system is a combination of DSS-7 (SBIG production) spectrometer and ST-402 (SBIG) CCD camera installed on Celestron 28cm (F/10) Schmidt-Cassegrain telescope. Also our system of multi-color photometry is a combination of ST-7E (SBIG) CCD camera accompanied with $B, V, R_c$ and Str\"{o}mgren $y$ filter attached to Celestron 23.5cm (F/6.3) Schmidt-Cassegrain telescope. The spectrometer's resolution $R = \lambda / \Delta\lambda$ is approximately 400 at 6000 \AA, and its dispersion is 5.4 \AA /pixel. 
Covering wavelength range is a 4200-8300 \AA. 

\section*{\normalsize 3. Results \& Discussion}

\hspace{0.5cm}  Figure 1 shows a result of our spectroscopic observations. 
We can see broad and prominent emission lines of Balmer series (H$\alpha$, H$\beta$, H$\gamma$), He I (4473, 5016, 5876, 7075), N II (4655, 5001, 5679), N III (4640), O I (7774) and Mg II (7880) on the spectra. Identification of these emission lines is based on those works by Anupama \& Dewangan (2000) $^6$ and Iijima (2002) $^7$. Numerical details of emission lines are shown in Table 1 and 2. The FWHM (Full Width at Half Maximum) of H$\alpha$ line is approximately 6200 km/s on Jan. 28.84UT, accompanying a blue shifted absorption line ($-4900$ km/s). On Jan. 29.84UT, H$\alpha$ line shows a nearly flat-topped profile in contrast to the previous day (see Figure 2), and the FWHM of H$\alpha$ has increased up to 7200 km/s. On the contrary, P Cygni profile is weakened. 
From these results we can see a rapid change of optical thickness. 
This is thought to be a reflection of the characteristics of this nova.
A general report of the spectroscopic observations in Japan is published separately 
(Yamanaka et al., 2010 $^8$; our data are partially included in this papaer.).
  
  We also performed multi-color photometry at the same time. Our data was processed by the differential (aperture) photometry using AIP4Win Version 2 (Berry \& Burnell, 2005 $^9$). Comparison star used is TYC6206-320-1 ($V=10.72$, $B-V=0.51$). The details of our results are shown in Table 3. In each band, about one magnitude decline per day is seen (rapid fading). Concerning the color indices, $B-V$ decreased but $V-R_c$ increased. 
According to van den Bergh \& Younger (1987) $^{10}$, color of the novae typically change bluer and our $B-V$ also shows smaller (bluer) during its decline. 
However our $V-R_c$ is not the case. 
Probably it is due to the $R_c$ band which contains broad H$\alpha$ emission line in this band.

\section*{\normalsize References}
\begin{enumerate}
\item Schaefer, B. E., 2010, \textit{ApJS}, \textbf{187}, 275
\item Pogson, N. R. et al., 1908, \textit{MRAS}, \textbf{58}, 90
\item Payne-Gaposchkin, C., 1957, \textit{The Galactic Novae} (North-Holland P.C.)
\item Maehara, H., 2010, \textit{vsnet-alert}, 11788
\item Schaefer, B. E., Harris, B. G., Dvorak, S., Templeton, M., and Linnolt, M., 2010, \textit{IAU Circ.} \textbf{9111}, 1S
\item Anupama, G. C. and Dewangan, G. C., 2000, \textit{AJ}, \textbf{119}, 1359
\item Iijima, T., 2002, \textit{A}\&\textit{A}, \textbf{387}, 1013
\item Yamanaka, M. et al., 2010, \textit{PASJ}, \textbf{62}, L37-L41
\item Berry, R. and Burnell, J., 2005, \textit{The handbook of astronomical image processing} (Willmann-Bell, Inc.)
\item van den Bergh, S. and Younger, P. F., 1987, \textit{A}\&\textit{AS}, \textbf{70}, 125
\end{enumerate}

\end{multicols}

\begin{figure}[H]
  \centering
  \includegraphics[width=14.5cm]{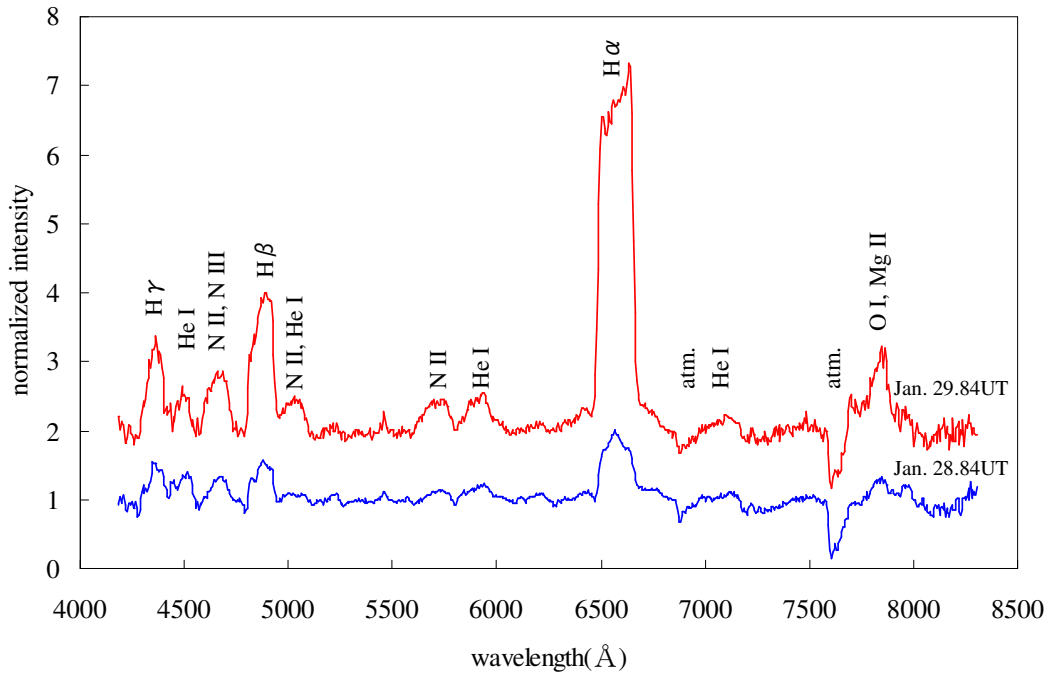}
  \caption{Spectra of U Sco on two nights. Lower spectrum was observed on Jan. 28.84UT, and upper spectrum was observed on Jan. 29.84UT. Each of these continuum is normalized as unity.} 
\end{figure}

\begin{figure}
  \centering
  \includegraphics[width=10cm]{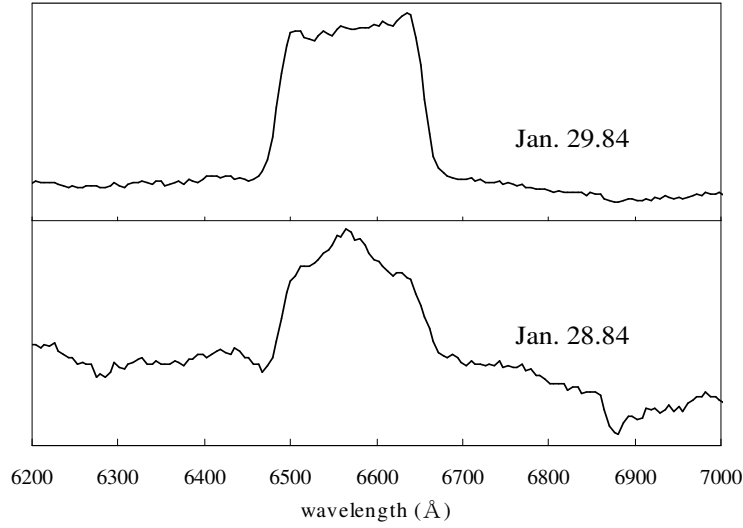}
  \caption{Spectra around H$\alpha$ line region on two nights. Upper spectrum shows a nearly flat-topped profile on Jan. 29.84UT compared with that on Jan. 28.84UT.} 
\end{figure}

\begin{table}
  \caption{Details of emission lines on Jan. 28.84UT. The \textit{P-Cyg} (km/s) is the blue shift of the center the P Cyg absorption component with respect to the center of emission component.}
  \centering
 \begin{tabular}{lccc} \hline
  Elements & $\lambda_{center} $(\AA) & FWHM (km/s) & \textit{P-Cyg} (km/s)\\ \hline
  H$\gamma$ (4340) & 4359 & 5400 & -5500\\
  He I (4473) & 4505 & 3400 & -\\
  N II (4655) \& N III (4640) & 4669 & 6200 & -6000\\
  H$\beta$ (4861) & 4879 & 6800 & -5300\\
  He I (5016) \& N II (5001) & 5033 & 7000 & -\\
  N II (5679) & 5718 & 6900 & -6100\\
  He I (5876) & 5909 & 7600 & -5600\\
  H$\alpha$ (6563) & 6576 & 6200 & -4900\\
  He I (7075) & 7117 & 3100 & -\\
  O I (7774) \& Mg II (7880) & 7837 & 3100 & -\\
  \hline
 \end{tabular}

  \caption{Details of emission lines on Jan. 29.84UT. Significant blue shift cannot be detected.}
  \centering
 \begin{tabular}{lccc} \hline
  Elements & $\lambda_{center} $(\AA) & FWHM (km/s) & \textit{P-Cyg} (km/s)\\ \hline
  H$\gamma$ (4340) & 4360 & 5600 & -\\
  He I (4473) & 4492 & 4300 & -\\
  N II (4655) \& N III (4640) & 4662 & 7500 & -\\
  H$\beta$ (4861) & 4875 & 8400 & -\\
  He I (5016) \& N II (5001) & 5035 & 6100 & -\\
  N II (5679) & 5702 & 8700 & -\\
  He I (5876) & 5914 & 6300 & -\\
  H$\alpha$ (6563) & 6576 & 7200 & -\\
  He I (7075) & 7101 & 5300 & -\\
  O I (7774) \& Mg II (7880) & 7841 & 3000 & -\\
  \hline
 \end{tabular}

  \caption{Results of multi-color photometry.}
  \centering
 \begin{tabular}{lccccccc} \hline
  date & JD & $B$ band & $V$ band & $R_c$ band & $y$ band & $B-V$ & $V-R$\\ \hline
  Jan. 28.84 & 2455225.34 & 8.88(6) & 8.60(4) & 8.07(2) & 8.66(7) & 0.28(10) & 0.53(6)\\
  Jan. 29.84 & 2455226.34 & 9.89(7) & 9.74(4) & 9.09(2) & 10.07(9) & 0.15(11) & 0.65(6)\\
  \hline
 \end{tabular}
\end{table}

\end{document}